\documentclass[a4paper,superscriptaddress,twocolumn,prd,numbers,showkeys]{revtex4}
\usepackage{newtxtext,newtxmath}
\usepackage{amsmath,mathrsfs,bm}
\usepackage{siunitx}
\usepackage{times}
\usepackage{graphicx}
\usepackage{subfigure,dcolumn}
\usepackage{booktabs}

\usepackage[colorlinks=true,linkcolor=blue,citecolor=blue,urlcolor=blue]{hyperref}
\usepackage{orcidlink}

\newcommand{\reffg}[1]{Figure~\ref{#1}}

\newcommand{\refeq}[1]{Equation~(\ref{#1})}
\newcommand{\refsc}[1]{Section~\ref{#1}}

\def\apjl{Astrophysical Journal Letters}
\def\aap{Astronomy and Astrophysics}

\def\apjs{The Astrophysical Journal Supplement Series}

\begin{document}

%\begin{CJK*}{GB}{}

\title{Configuration Requirements for 21-cm Forest Background Quasar Searches with the Moon-based Interferometer}
\author{Siyuan Zhang \,\orcidlink{0009-0000-1677-0737}}
\affiliation{Key Laboratory of Cosmology and Astrophysics (Liaoning), College of Sciences, Northeastern University, Shenyang 110819, China}

\author{Qi Niu \,\orcidlink{0009-0007-1168-0928}}
\affiliation{Key Laboratory of Cosmology and Astrophysics (Liaoning), College of Sciences, Northeastern University, Shenyang 110819, China}

\author{Yichao Li\,\orcidlink{0000-0003-1962-2013}}
% \correspondingauthor{Yichao Li}
\email[Corresponding author, ]{liyichao@mail.neu.edu.cn}
\affiliation{Key Laboratory of Cosmology and Astrophysics (Liaoning), College of Sciences, Northeastern University, Shenyang 110819, China}

\author{Xin Zhang\,\orcidlink{0000-0002-6029-1933}}
% \correspondingauthor{Xin Zhang}
\email[Corresponding author, ]{zhangxin@mail.neu.edu.cn}
\affiliation{Key Laboratory of Cosmology and Astrophysics (Liaoning), College of Sciences, Northeastern University, Shenyang 110819, China}
\affiliation{National Frontiers Science Center for Industrial Intelligence and Systems Optimization, Northeastern University, Shenyang 110819, China}
\affiliation{Key Laboratory of Data Analytics and Optimization for Smart Industry (Ministry of Education), Northeastern University, Shenyang 110819, China}

%% Mark off the abstract in the ``abstract'' environment. 
\begin{abstract}
The 21-cm forest offers a powerful cosmological probe of the thermal history and small-scale structure of the intergalactic medium during the Epoch of Reionization (EoR). Its success, however, critically depends on the availability of high-redshift radio-loud quasars (HzRLQs) as background sources. In this work, we investigate the configuration requirements for a Moon-based low-frequency radio interferometer aimed at maximizing the detection of HzRLQs for future 21-cm forest studies. Building upon a previously developed quasar luminosity function (QLF), we forecast HzRLQ abundances under various array configurations. Assuming a total survey area of $10^4\,\mathrm{deg}^2$ and 1 year of observation, we compare continuum surveys with 10 MHz bandwidth and 21-cm forest surveys with 5 kHz resolution. Our results show that a minimum collecting area of $\sim$6500 m$^2$ enables detection at $z \sim 6$, while SKA-like arrays ($N_{\mathrm{st}} = 512$) extend the detection limit to $z \sim 10$ for 21-cm forest survey and $z \sim 16$ for continuum survey. Larger arrays with $N_{\mathrm{st}} = 2048$ can reach $z \sim 11$ in 21-cm forest mode. We also explore configurations that maintain fixed collecting areas while increasing the number to enhance survey efficiency. This boosts source detection but significantly increases the data volume and computational demands. These results underscore the importance of optimizing array design for different survey goals and balancing sensitivity, spectral resolution, and data management. A well-designed Moon-based array could open a new observational window on reionization and early cosmic structure formation.
\end{abstract}

%\date{Accepted XXX. Received YYY; in original form ZZZ}
%% Keywords should appear after the \end{abstract} command. 
%% The AAS Journals now uses Unified Astronomy Thesaurus concepts:
%% https://astrothesaurus.org
%% You will be asked to selected these concepts during the submission process
%% but this old "keyword" functionality is maintained in case authors want
%% to include these concepts in their preprints.

\keywords{ Quasars  --- Radio loud quasars  --- Reionization  --- HI line emission  --- Surveys }

\maketitle

\section{Introduction} \label{sec:intro}

The 21-cm signal from the hyperfine transition of neutral hydrogen (HI) has become 
a powerful cosmological probe for the study of the early Universe, in particular, 
the Dark Ages, the Cosmic Dawn, and the Epoch of Reionization (EoR) 
\citep{2008PhRvD..78j3511P}. 
The 21-cm absorption features imprinted by HI in the early Universe on the spectra of high-redshift radio-loud background sources has emerged as a promising probe of the intergalactic medium (IGM) during the Cosmic Dawn and EoR \citep{2002ApJ...577...22C,2014PhRvD..90h3003S,2020ApJ...899...16T}.
Recently, \citet{2023NatAs...7.1116S} proposed a novel cosmological probe known as 
the 21-cm forest, which provides a unique probe of small-scale structures during the EoR. 
The 21-cm forest relies on the absorption characteristics of HI in the 
intervening structures along the line of sight with respect to the spectra of the 
bright background radio sources, such as radio-loud quasars. 
This technique provides valuable insights into the nature of dark matter (DM) and its 
role in shaping small-scale structure formation
\citep{2016JCAP...08..004L,2006PhR...433..181F,2012RPPh...75h6901P,2025PhLB..86239342S,2025arXiv250414656S}.

%Despite its potential, 
Breakthroughs in the 21-cm forest have been difficult to achieve because of ongoing challenges 
in both analytical modeling and observations.
%, mainly due to some limiting factors. 
Because of the absence of analytical models, the parameter inference of the 21-cm forest relies on complex small-scale simulations\citep{2021MNRAS.506.5818S,2023MNRAS.519.3027S}. 
The substantial computational cost of simulations poses a significant challenge to constrain the astrophysical 
parameters.
Recently, several new approaches have been proposed to address this challenge, including the deep-learning-driven likelihood-free parameter inference methods \citep{2024arXiv240714298S} and a halo-model-based analytical model for the one-dimensional power spectrum of the 21-cm forest \citep{2024arXiv241117094S}. %Secondly, 21-cm forest observations require HzRLQs as background sources. The number of known HzRLQs remains limited, highlighting the urgent need for future surveys to discover more suitable candidates.

Moreover, the detection of the 21-cm forest power spectrum remains challenging due to the limited
sensitivity of current low-frequency observational facilities. 
This constraint is especially critical for the identification of high-redshift radio-loud quasars (HzRLQs),
which serve as essential background sources.
While optical and near-infrared surveys have extended quasar observations to 
redshifts beyond 6 \citep{2015Natur.518..512W,2023Natur.621...51D,2022A&A...662L...2Z}, the number of detected radio-loud 
quasars at such high redshifts remains small. 
A recent study \citep{2025ApJ...978..145N} examined the abundance of HzRLQs 
and confirmed that the observations are still constrained by flux limitations
even with the next-generation low-frequency radio telescope arrays, such as the 
low-frequency Square Kilometre Array (SKA-Low). 
This scarcity presents a significant obstacle to advancing 21-cm forest studies.
%because detecting individual absorption lines requires bright and distant radio sources.

Advances in space technology now make it possible to utilize the far side of the Moon $-$ a radio-quiet environment that is uniquely shielded from Earth's ionospheric aberrations and man-made radio frequency interference (RFI)\citep{2009SPIE.7436E..0IL,2010PhRvD..81d2003J,2018SPIE10704E..2SI}. This pristine electromagnetic environment provides an extraordinary platform for groundbreaking low-frequency radio astronomy observations \citep{2012P&SS...74..167K}. The RFI-free conditions on the far side of the Moon allow the detection of weak cosmological signals, in particular, the 21-cm HI line from the Dark Ages, Cosmic Dawn, and EoR. %that is simply undetectable from Earth
\citep{2021arXiv210308623B,2024AdSpR..74..528P}. 

In this work, we thoroughly explore the survey strategies for HzRLQs with the future moon-based low-frequency array.
Although the technology for building a large radio interferometric array like SKA-Low 
on the far side of the Moon remains underdeveloped, this paper explores its immense potential and the fundamental configuration requirements for detecting HzRLQs. Specifically, we analyze two deployment scenarios: one involving variations in the effective receiving area and the other involving changes in the station diameter. These approaches are designed to address existing observational limitations. Additionally, we provide a detailed comparison of the HzRLQs under both the continuum survey and the 21-cm forest survey. However, these strategies also introduce new engineering challenges related to array optimization and data processing. Such a Moon-based observatory will provide transformative insights into the formation of the cosmic structure at the dawn of the universe and during its first billion years \citep{1990AIPC..207..522K, 2024RSPTA.38230094C,2024SPIE13092E..2LI}.

The paper is organized as follows: In Section \ref{sec:2}, we describe the construction of the luminosity function for predicting the abundance of HzRLQs and the improvement of the model by including the obscuration effects. 
In Section \ref{sec:conf}, we present various scenarios for the construction of the low-frequency radio array on the Moon. 
In Section \ref{sec:3}, we present our results, including the influence of obscuration effects on the prediction of high-redshift quasars, how to build future moon-based arrays with the most significant detection of quasars and cost savings, and the baseline requirements to minimize the confusion effects. Finally, in Section \ref{sec:4}, we summarize our conclusions and outline future research directions.

\section{Luminosity function model}\label{sec:2}

\subsection{Physical-driven model}

This study extends the physical-driven model established in previous work by incorporating the impact of optical observation biases on the detection of obscured quasars. A brief summary of the model is provided here, with detailed descriptions available in \cite{2025ApJ...978..145N}. The key steps in constructing the model are described as follows:

\begin{enumerate}
\item  \textbf{Calculating the black hole (BH) mass function.} Quasars are driven by the active behavior of BHs at the centers of galaxies. Therefore, the first step is to determine the abundance of BHs in galaxy centers. We use a dark matter halo mass function in Sheth--Tormen form \citep{2002MNRAS.329...61S} to describe the abundance of halos of different masses at different redshifts. We assume that each halo hosts a galaxy and that each galaxy contains a central BH. The BH mass function is then derived from the halo mass function using the mass relationship between BHs ($M_{\rm BH}$) and their corresponding halos ($M_{\rm h}$) \citep{2003ApJ...595..614W,2025ApJ...978..145N},
\begin{align}
M_{\rm BH}(M_{\rm h}, z) = A \left(\frac{M_{\rm h}}{10^{12}M_{\odot}}\right)^{\frac{5}{3}}
\left(\frac{\xi(z)(1+z)^3}{\Omega_{\rm m}}\right)^{\frac{5}{6}} M_{\odot},
\end{align}
where $A$ is the amplitude parameter and $\xi(z)$ is the dimensionless 
parameter related to redshift $z$. In this work, we adopt the best-fit value $A \simeq 3.2\times10^{6}$ from \cite{2025ApJ...978..145N}.

\item  \textbf{Determining the quasar duty cycle with quasar luminosity function (QLF).} The quasar duty cycle $D_{\rm q} = t_{\rm q}/t_{H(z)}$ quantifies the fraction of black holes in the quasar phase, where $t_{\rm q}$ is the quasar lifetime and $t_{H(z)}$ is the cosmic age at redshift $z$.
Assuming Eddington-limited accretion, we constrain $t_{\rm q}$ by fitting observational data of the QLF at high redshifts \citep{2018ApJ...869..150M,2023ApJ...949L..42M}.

\item  \textbf{ Determining the fraction of radio-loud quasars whose radio flux exceeds the required threshold $F_{th}$. } 
We assume that $10\%$ of quasars are radio-loud 
\citep[according to observational experience, e.g.][]{2000AJ....119.1526S,2015ApJ...804..118B,2021ApJ...908..124L}. To account for the radio flux distribution of the quasar population at a given optical luminosity, we adopt the observed distribution of radio-loudness $R \equiv {\rm log_{10}}\left(L_{\rm 5GHz} / L_{4400}\right)$, where $L_{\rm 5GHz}$ and $L_{4400}$ are the 
luminosity measured at radio band ($5\, {\rm GHz}$) and optical band ($4400\,{\rm \AA}$), respectively. 
We use a radio-loudness distribution function $N(R)$ constrained using the observation 
data from \citet{2022MNRAS.513.4673B}.
The fraction of radio-loud quasars over a flux threshold $F_{\rm th}$ is determined by 
integrating the radio-loudness distribution function $N(R)$,
\begin{align}
\epsilon (M_{\rm h}, F_{\rm th}) = \int_{R_0}^{\infty} {\rm d}R N(R),
\end{align}
where the lower-bound $R_0 = R_0(M_{\rm BH}(M_{\rm h}, z), F_{\rm th}) $ 
is a function of black hole mass and flux threshold. 

\item  \textbf{Estimate the abundance of HzRLQs.} 
The abundance of HzRLQs is then estimated by integrating the weighted halo mass function,
\begin{align}
N(&\Delta z, \Delta \Omega, F_{\rm th}) \nonumber \\ 
= &\int_{\Delta z, \Delta \Omega} {\rm d} V \int {\rm d} M_{\rm h} 
10\% D_{\rm q} \epsilon(M_{\rm h}, F_{\rm th}) \Phi_{\rm h}(M_{\rm h}),
\end{align}
where $\Delta z$ and $\Delta \Omega$ are the redshift bin and survey sky area, 
${\rm d}V$ is the cosmological volume element, 
and $\Phi_{\rm h}(M_{\rm h}) = {\rm d}n/{\rm d}M_{\rm h}$ represents the halo mass function.

\end{enumerate}

\subsection{Obscuration-corrected QLF}

%We find that a key factor that may affect the prediction of the abundance of high-redshift quasars is obscuration. 
%The obscuration in the optical band may affect the prediction of the abundance of the HzRLQs.
%The distribution and nature of dust have significant impacts on the observation of AGN. Especially in molecular clouds and regions with high extinction, the shading effect of dust may make AGN invisible in the optical band \citep{Fanciullo_2017}. In the optical band, quasars can be heavily obscured by dust, making them difficult to detect. This obscuration effect has been well documented in studies of active galactic nuclei (AGNs), where a significant fraction of AGNs are missed in optical surveys due to extinction by dust \citep{Guo_2021}, suggesting that current predictions of the abundance of high-redshift quasars, which are based primarily on optical observations, may be underestimating the true number of quasars that are available to the 21-cm forest study.

The obscuration in the optical band may significantly affect predictions of 
the abundance of the HzRLQs. 
The distribution and nature of dust play a critical role in the 
observational properties of active galactic nuclei (AGNs). 
In particular, the molecular clouds and high-extinction regions can obscure AGNs,
rendering them undetectable in optical surveys \citep{2017A&A...602A...7F}. 
Quasars, especially at high redshifts, can be heavily obscured by dust in the optical band, 
which complicates their identification. 
This obscuration effect has been extensively studied, with findings indicating that
a considerable fraction of AGNs are missed in optical surveys due to dust extinction \citep{2021ApJ...906...47G}. 
%As a result, current estimates of high-redshift quasar abundance, 
%primarily based on optical observations, may be significantly underestimated. 
Since the construction of the HzRLQ abundance model depends on the QLF in the optical band, 
dust obscuration in the optical band may lead to an underestimation of the abundance of HzRLQs. 
This underestimation has direct implications for studies such as the 21-cm forest, 
which rely on a complete census of background radio sources.
%In heavily obscured environments, such as molecular clouds and high-extinction regions, 
%dust can significantly attenuate optical emission, rendering some AGN undetectable \citep{Fanciullo_2017}.
%\citet{Guo_2021} further demonstrates that a substantial fraction of AGNs are missed 
%in optical surveys due to dust extinction.

%In a high-redshift cosmic environment, dense gas clouds can obscure quasars in the background, and this obscuration is achieved mainly by absorbing the light emitted by the quasar. 
The obscuration renders the quasar luminosity in certain wavelength bands. 
To ensure the accuracy of the observed high-redshift quasar population, their 
luminosity function can be corrected by introducing an obscuration fraction, 
which is estimated based on the observations in the X-ray band \citep{2023ApJ...950...85L}.
\begin{equation}
    f_{\rm obsc}={\rm min}[\psi_{\rm max},{\rm max}[\psi_0-\beta \log(L_{\rm X}/L_{{\rm X},0}),\psi_{\rm min}]],
\end{equation}
%The $f_{\rm ob}$ parameter represents the fraction of obscured AGN, 
where $\phi_0=0.73$, $\psi_{\rm max} = 0.84$, $\psi_{\rm min}=0.2$, $\beta = 0.24$, 
$L_{\rm X,0}=10^{43.75}{\rm erg\,s}^{-1}$ \citep{2011PASJ...63S.937U,2014ApJ...786..104U}, 
%are the maximum and minimum fractions of obscured AGNs found in the highest luminosity range of the Swift/BAT survey, respectively. 
%We used the best-fit values obtained in \cite{2011PASJ...63S.937U} to make $\beta=0.24$, and due to the limited number of sources, we only include $\psi_0$ as a free parameter, which represents the fraction of obscured AGNs among all AGNs at  \citep{Ueda_2014}.
The X-band luminosity is estimated via the bolometric correction factor of the hard X-ray band
\cite{2020A&A...636A..73D}:
\begin{equation}
    K_{\rm X}(L_{\rm bol}) = \frac{L_{\rm bol}}{L_{\rm X}} 
    = a\left( 1+\left(\frac{{\rm log}(L_{\rm bol}/L_\odot)}{b}\right)^c\right)
\end{equation}

where $a=10.96$, $b=11.93$, and $c=17.79$.
It is worth noting that the obscuration fraction function is primarily constrained
using X-ray selected AGN samples at redshifts $z \lesssim 5$. 
Due to the limited observational data available for quasars at higher redshifts, 
the redshift evolution of the obscuration fraction remains uncertain. 
In this work, we therefore adopt a simplified approach and assume a 
redshift-independent obscuration fraction, 
acknowledging that this may introduce some uncertainties in our predictions.

\begin{figure}
    \centering
    \includegraphics[width=0.48\textwidth]{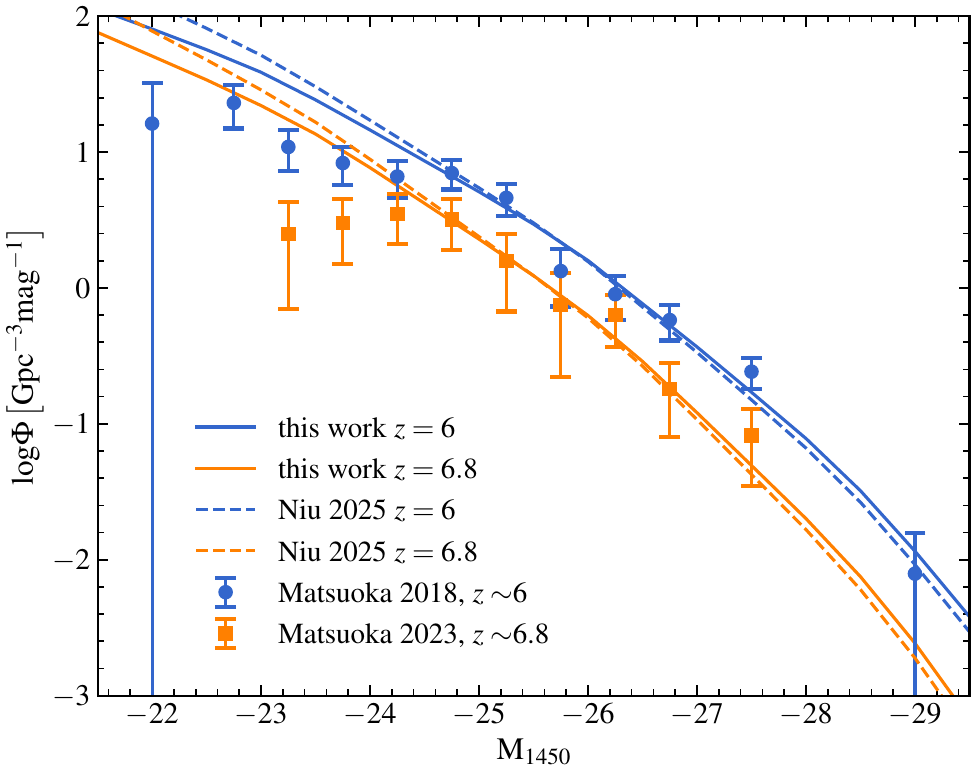}
    \caption{
    Comparison of the best-fit QLF with and without considering the optical-band
    obscuration effect. 
    Blue and orange curves denote models at redshifts $z=6$ and $z=6.8$, respectively. 
    Solid lines illustrate models with obscuration effects in our work, 
    whereas the dotted lines represent models without considering the obscuration effect. 
    The blue and orange errorbars show the measurements at $z\sim 6$ from \citet{2018ApJ...869..150M}
    and $z\sim6.8$ from \citet{2023ApJ...949L..42M}.
    Owing to observational incompleteness, the first two data points on the faint end are omitted.}\label{fig:1}
\end{figure}

% We utilize the model of the luminosity function that can be applied to predict high-redshift quasars, where the thermal luminosity is expressed in terms of the Eddington luminosity. The $M_{1450}$ is denoted as
% \begin{equation}
%     M_{1450}=-2.5\rm log(\frac{L_{\nu}(\nu(1450\rm \mathring{A}))}{4\pi d^23631\rm Jy})
% \end{equation}
%  where d = 10 pc is expressed in centimeters.
 
We correct the luminosity function by multiplying the luminosity function of total quasars
with the fraction of unobscurated to obtain the luminosity function of the optically observed quasars,
\begin{equation}
    %\Phi_{\rm unob}=\Phi(M_{1450},z) \times (1-f_{\rm ob}).
    \Phi = \Phi_{\rm total}(M_{1450},z) \times (1-f_{\rm obsc}).
\end{equation}

%where $\Phi(M_{1450},z)$ corresponding to the dotted lines in \reffg{fig:1} is the case where the obscuration effect is not taken into account, while $\Phi_{\rm unob}$ corresponding to the solid lines is the luminosity function of the unobserved portion taking into account the obscuration effect.

% In the derivation of the Quasar Luminosity Function (QLF), the parameter $t_{\rm q}$ (the mean lifetime of the quasar) is not strictly constrained by observational data. Since these parameters together shape the QLF, they show a strong correlation when fitting the QLF to the observational data set. In logarithmic space, based on a discussion of accretion rates, we optimize the fit of the QLF to the observed data by adjusting the mean lifetime $t_{\rm q}$ until the chi-square ($\chi^2$) statistic reaches a minimum. This approach yields the best fit to the observed luminosity function data. We utilized two datasets: one observed at $z\sim6$ \citep{2018ApJ...869..150M} and the other at $z\sim6.8$ \citep{2023ApJ...949L..42M}. Our best-fit results are shown in \reffg{fig:1}. For the fits at z = 6 and z = 6.8, the smallest odd-squared values are about 7.79 and 3.78, respectively.We then used the logarithmic mean of these values for further calculations.

We compare the best-fit model with and without considering the obscuration effect in \reffg{fig:1}.
The QLF model with considering the obscuration effect, is shown with the solid curve. 
The dashed curve represents the best-fit model from \citet{2025ApJ...978..145N}.
%We discuss in logarithmic space the accretion rate by 
The QLF model is fitted to the data by  adjusting the mean quasar lifetime $t_{\rm q}$ 
until the likelihood function is maximized. 
Two main datasets are used in this work, i.e., one set of data is at $z\sim 6$
\citep{2018ApJ...869..150M}, and the other one is at $z\sim 6.8$ \citep{2023ApJ...949L..42M}. 
For the fits at $z = 6$ and $z = 6.8$, the least-square values are about $7.79$ and $3.78$,
respectively. 
Taking into account the obscuration effect, the luminosity function 
has $t_{\rm q}=10^{5.79}$ at $z=6$, $t_{\rm q}=10^{5.76}$ at $z=6.8$.

%After the introduction of the obscured effect, the function to be fitted changes from $\Phi(M_{1450},z)$ to $\Phi_{\rm unob}$, and the obscured effect mainly affects the dark-end data \citep{Ueda_2014}, so the overall slope of the new fit slows down compared to the original one.

% For the fits at $z=6$ and $z=6.8$, the reduced chi-square values were approximately and, respectively. While the fit appears to improve with a higher Eddington ratio, it is crucial to acknowledge that this result may still be influenced by the incomplete nature of the high-redshift QLF observations. This incompleteness can be attributed to two main factors. First, the sensitivity limitations lead to missing data at the faint end of the QLF, which we addressed by discarding the relevant data points. Second, the completeness of optical observations may be compromised by the obscuration from AGN dust tori, rendering some AGNs undetectable in the optical band.

By comparing the results with and without obscuration effects,
we find that incorporating obscuration leads to a slightly flatter QLF, 
a trend observed at both redshifts considered in our analysis. 
This is mainly because low-luminosity quasars are more likely to be obscured by dust, 
making them less likely to be detected in optical surveys. 
Including the obscuration effect can thus partially alleviate the discrepancy 
between theoretical models and observational QLF data at the faint end. 
However, the improvement is limited. As mentioned above, the obscuration fraction 
is based on low-redshift data, and its possible redshift evolution is not considered in this work. 
Additionally, the incompleteness of the observed sample at the faint end may 
still be the dominant factor contributing to the observed deviation.

%\subsection{Sensitivity estimation}
%
%%To predict the number of high-redshift observations for SKA-Low, we use the following form of noise.
%
%%For continuous measurements, 
%%the field of view of a single field \( \Omega \) is given by \( W \sim \left(\frac{1.2\lambda} {D_{\text{ant}}}\right)^2 \), where \( D_{\text{ant}}\) is the diameter of the station. 
%
%
%In this work, we combine the obscuration-corrected optical luminosity function with the radio luminosity distribution to compute the radio luminosity function. 

\section{Moon-based interferometer configuration}\label{sec:conf}

Literature research indicates that a significant detection of HzRLQs requires a radio interferometer with a 
sensitivity comparable to that of SKA-Low \citep{2025ApJ...978..145N}.
SKA-Low is designed to probe the early Universe through radio observations. 
It consists of $131072$ log-periodic dipole antennas, grouped into $512$ stations, 
each containing $256$ antennas 
\footnote{\url{https://www.skao.int/en/explore/telescopes/SKA-Low}}.
Approximately $50\%$ of these stations are concentrated within a $1\, {\rm km}$ diameter central core, 
while the rest are distributed along three spiral arms extending up to $74\, {\rm km}$. 
The total effective collecting area is about $419000\,{\rm m}^2$, providing exceptional sensitivity to faint radio signals. 
Operating within a frequency range of $50\, {\rm MHz}$ to $350\, {\rm MHz}$, 
SKA-Low can observe the 21-cm signal at redshifts approximately between $z \sim 3$ and $z \sim 27$, 
covering key cosmic epochs such as the EoR and the Cosmic Dawn. 

The observational noise variance of the radio interferometer is estimated via,
\begin{equation}
    \sigma=\sqrt{2}\frac{k_{\rm B}T_{\rm sys}}{A_{\rm eff}}
    \frac{1}{\sqrt{N_{\rm st}(N_{\rm st}-1)\Delta t\Delta \nu}},
\end{equation}
where $k_{\rm B}$ is the Boltzmann constant, 
$A_{\rm eff}$ refers to the effective receiving area of a station, 
$N_{\rm st}$ is the number of stations, 
$\Delta\nu$ is the frequency channel width, and $\Delta t$ is the integration time. 
$T_{\rm sys}$ is the system temperature, and it is consist of,
\begin{equation}
T_{\rm sys} = T_{\rm sky} + T_{\rm rx}
\label{eq:8}
\end{equation}
where $T_{\rm sky}$ is the mean brightness temperature from the sky and takes the form of
\begin{equation}
T_{\rm sky} = \left(2.73 + 25.2 \times (408/\nu_{\rm MHz}(z))^{2.75}\right)\,{\rm K}, 
\end{equation}
where $\nu(z)$ represents the frequency in $\rm MHz$ corresponding to redshift $z$. 

In this work, we consider that the surveys are carried out with a continuum survey, which has the flux measurements integrated 
over the frequency band of $10\,{\rm MHz}$ to enhance the signal-to-noise ratio for continuum source detection. An integration bandwidth of 10 MHz was selected to achieve an optimal balance between resolving the continuum's spectral shape and maintaining sufficient detection sensitivity. This bandwidth is sufficiently narrow to capture spectral features while remaining broad enough to minimize noise.
Due to the limited frequency resolution, the following 21-cm forest analysis requires additional 
deep-field observations. We also consider a 21-cm forest survey with spectrum resolution of $5\,{\rm kHz}$. 
Such high-frequency resolution significantly reduces the detection ability for HzRLQs, but it is fine enough for 
the 21-cm forest analysis without additional follow-up observations.

We assume $T_{\rm rx} = 0.1T_{\rm sky}+40\,{\rm K}$ \citep{2020PASA...37....7S} and also assume a constant total survey area $S_{\rm tot} = 10000\,{\rm deg^2}$
and total observation time $t_{\rm tot} = 1\,{\rm yr}$. The observational noise variance may vary according to 
different effective collecting areas and survey efficiency. 
It is worth noticing that the sky temperature $T_{\rm sky}$ dominates the noise term at the low frequency bands, as shown in \refeq{eq:8}. We systematically varied the receiver temperature to $T_{\rm rx} = 0.1T_{\rm sky}+70\,{\rm K}$ and find that it is only $6.37\%$ more than the original. This analysis demonstrates that variations in receiver noise temperature have a negligible influence on the total system noise under these conditions.
%In order to better realize the construction of the SKA-Low radio telescope on the Moon, we will explore the minimum construction size of the SKA to satisfy that at least 10 bright quasars can be observed at various redshifts. We also vary the size of the individual antennas and hence the number of base stations to achieve a larger field-of-view survey while keeping the effective receiving area constant.

\paragraph{Varying the effective collecting area}
We consider the future construction of low-frequency interferometric arrays of varying sizes on the far side of the Moon. 
Using SKA-Low as a reference, we maintain a constant station diameter of $D = 40\,{\rm m}$. 
The array size ranges from $N_{\rm st} = 8$, $32$, $128$, and $512$ up to $2048$ stations. 
Notably, an array with $512$ stations is comparable in size to the current SKA-Low, 
while the 2048-station configuration represents a more ambitious scenario, 
with an effective collecting area exceeding one square kilometer.

%We made several attempts to change the size of the effective receptive area while the antenna length was constant at $40\, {\rm m}$ to achieve the sighting of 10 effective quasars at z= 6, 8, 10, and 12. This can lead to the conclusion that SKA-Low should be constructed at least as large as possible to meet different observation needs during lunar construction.

\paragraph{Varying the station diameter}
The solid angle of the field of view (FoV) for a single pointing is estimated using the primary beam size of the station
given by $\Omega \sim (1.2 \lambda / D_{\rm st})^2$, where $D_{\rm st}$ is the diameter of a station.
For the subsequent analysis, we use the wavelength $\lambda$ corresponding to $150$ MHz to estimate the FoV.
By varying the station diameter, the FoV for a single pointing changes, thereby affecting the survey efficiency, 
while keeping the total survey area and observation time constant. 
To maintain consistency with the current SKA-Low, we ensure the total effective collecting area remains at $\sim 419000\,{\rm m}^2$,
and varying the number of stations and the diameter of the station simultaneously.
In particular, we consider the following station configurations:
$\{(N_{\rm st}, D_{\rm st})\} = \{(512, 40\,{\rm m}),  (2048, 20\,{\rm m}),  (8192, 10\,{\rm m})\}$.

%We know that the size of a base station, i.e., the effective area of a single base station, is directly proportional to the number of antennas and directly proportional to the square of the length of the antennas. If we want to keep the total effective area constant and reduce the area of individual base stations, we have to increase the number of base stations.

%On the premise of keeping the projected effective receiving area of $41000 \rm m^2$  for the SKA-Low project unchanged, we reduced the length of individual antennas from the original size to 10 meters and increased the number of base stations accordingly. In this way, although the receiving area of a single base station is reduced, the field of view area for continuous sky patrol is significantly increased. On this basis, we continuously observe a single deep-field area for 40 hours and compare and analyze the observed noise signal strength.

\section{Results and Discussions}\label{sec:3}

Using the configuration parameters introduced in \refsc{sec:conf}, we estimate the system noise variance level
for each case. We use $10$ times of the noise variance level as the lower bound of the integration 
of the predicted luminosity function to forecast the abundance of HzRLQs.

\subsection{Configuration requirements with continuum survey}

\begin{figure*}
    \centering
    \begin{minipage}[t]{0.48 \textwidth}
    \includegraphics[width=\textwidth]{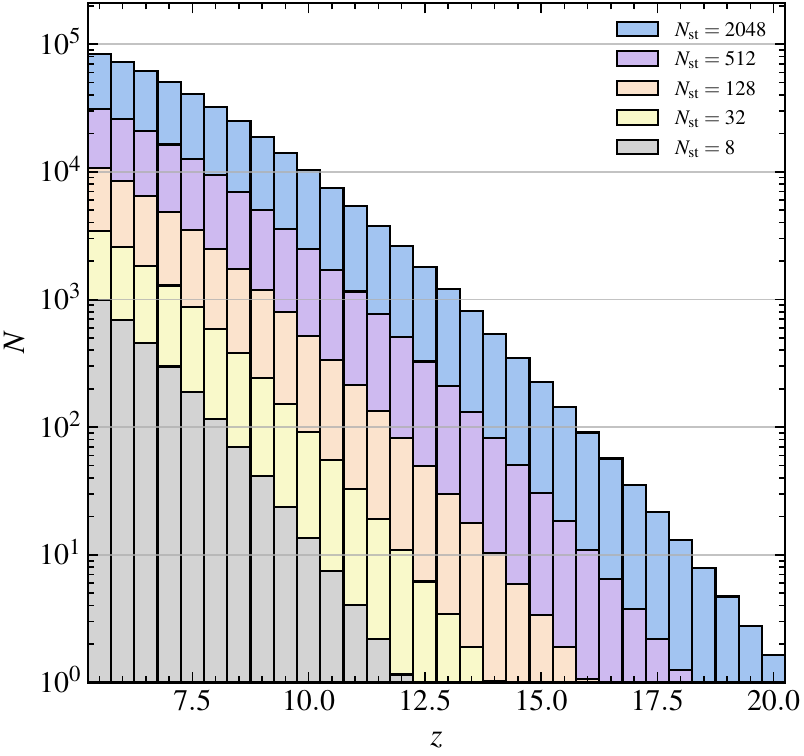}
    \caption{The number of HzRLQs detected with continuum survey as a function of redshift with different sizes of interferometer.
    The results with $N_{\rm st} = 8$, $32$, $128$, and $512$ up to $2048$ stations are shown in different colors.
    %The figure shows the variation of the number of quasars with redshift observed by the SKA-Low array Continuum Survey with different numbers of base stations.The bars of different colors represent the observations of SKA arrays with different numbers of base stations.
    }\label{fig:2}
    \end{minipage} \hfill
%\end{figure}
%\begin{figure}
%    \centering
    \begin{minipage}[t]{0.48 \textwidth}
    \includegraphics[width=\textwidth]{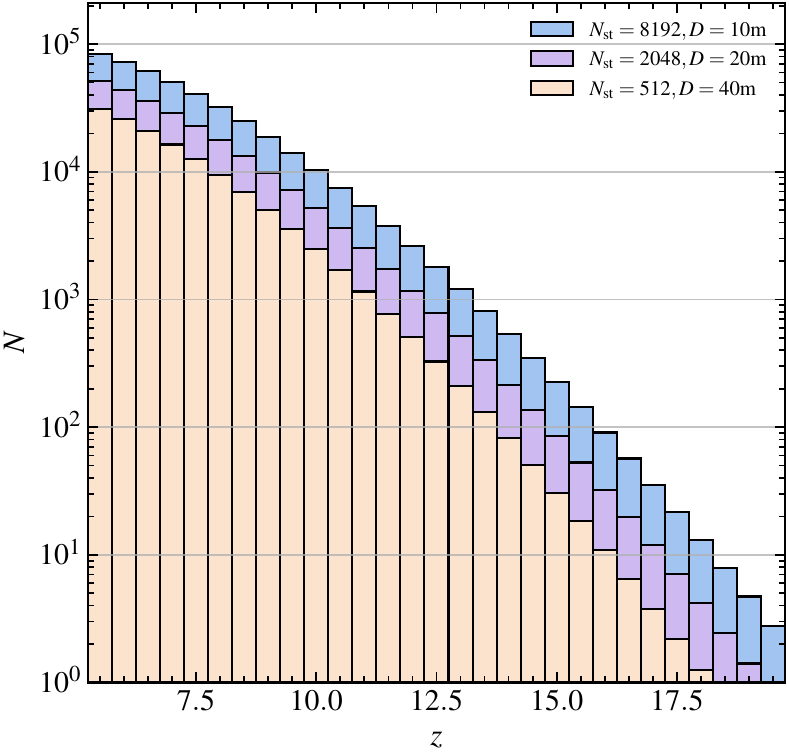}
    \caption{The number of HzRLQs detected with continuum survey as a function of redshift with different diameters of station and
    number of station configurations.
    The results with $\{(N_{\rm st}, D_{\rm st})\} = \{(512, 40\,{\rm m}),  (2048, 20\,{\rm m}),  (8192, 10\,{\rm m})\}$ 
    are shown in different colors.
    %Bars of different colors represent the variation with redshift in the number of quasars that can be observed with redshift at successive sky surveys for the number of base stations constructed at different antenna lengths.
    }\label{fig:4}
    \end{minipage}
\end{figure*}

The number of HzRLQs detected (with their flux over 10 times the corresponding noise level) in continuum surveys 
is shown in \reffg{fig:2}, where different colors represent results for various interferometer size configurations. 
Since the predicted abundance depends on the width of the redshift bins, we fix the redshift bin $\Delta z = 0.5$ 
for the analysis of continuum surveys.
Generally, the number of detected HzRLQs significantly reduced as redshift increased. 
The gray histogram corresponds to an array of eight $40\,{\rm m}$-diameter stations, 
with an effective collecting area of $6546.875\,{\rm m}^2$. 
This configuration enables the detection of approximately $10^3$ radio-loud quasars 
at $z \sim 5$ and provides statistically significant detections ($N > 10$) up to $z \sim 10$.

Increasing the number of stations dramatically enhances detection capabilities. 
With $N_{\rm st} = 128$, the number of detected HzRLQs increases by an order of magnitude at each redshift bin, 
and the redshift detection limit extends to $z \sim 16$.

As discussed in \refsc{sec:conf}, an interferometer with 512 $40\,{\rm m}$-diameter stations achieves
an effective collecting area comparable to that of the current SKA-Low.
An even more ambitious configuration, with $N_{\rm st} = 2048$, 
would push the total collecting area beyond one square kilometer. 
Such an interferometer would enable the robust detection of HzRLQs up to the 
beginning of the EoR or even Cosmic Dawn.

Increasing the effective collecting area of an interferometric array 
can significantly enhance its sensitivity to HzRLQs. 
Beyond simply scaling up the array, we also explore an alternative strategy 
that maintains the same total collecting area -- comparable to that of the SKA -- 
while boosting survey efficiency by employing smaller station diameters to achieve a larger field of view. 
The corresponding results are shown in \reffg{fig:4}, where different colors indicate 
configurations with varying station numbers and sizes.

Our findings demonstrate that even without increasing the total collecting area,
the detection capability for HzRLQs can be improved by increasing the number of stations 
and reducing the diameter of each station. 
This configuration effectively enhances sky coverage and survey speed. 
However, it is important to note that such a design imposes substantial demands on the system, 
particularly in terms of data transmission, communication bandwidth, 
and computational resources required for data processing. 
These factors pose significant challenges for the data analysis pipelines of interferometric arrays 
and must be carefully considered in future instrument design.

Our analysis shows that the detectability of HzRLQs in continuum surveys 
is strongly dependent on both the size and configuration of the interferometric array. 
Increasing the number of stations significantly boosts sensitivity, 
enabling detections out to higher redshifts, with large arrays ($N_{\rm st} \gtrsim 512$) 
capable of probing the EoR and potentially even Cosmic Dawn. 
Moreover, we demonstrate that optimizing array layout 
-- by increasing station number while reducing individual station size -- 
can further enhance survey efficiency without increasing the total collecting area. 
This approach increases the survey field of view and improves detection rates but introduces substantial 
challenges in data transmission and processing. 
These trade-offs highlight the need for carefully balanced array designs that optimize both scientific return and technical feasibility in future low-frequency radio surveys targeting HzRLQs.

\subsection{Configuration requirements with 21-cm forest survey}

\begin{figure*}
    \centering
    \begin{minipage}[t]{0.48\textwidth}
    \includegraphics[width=\textwidth]{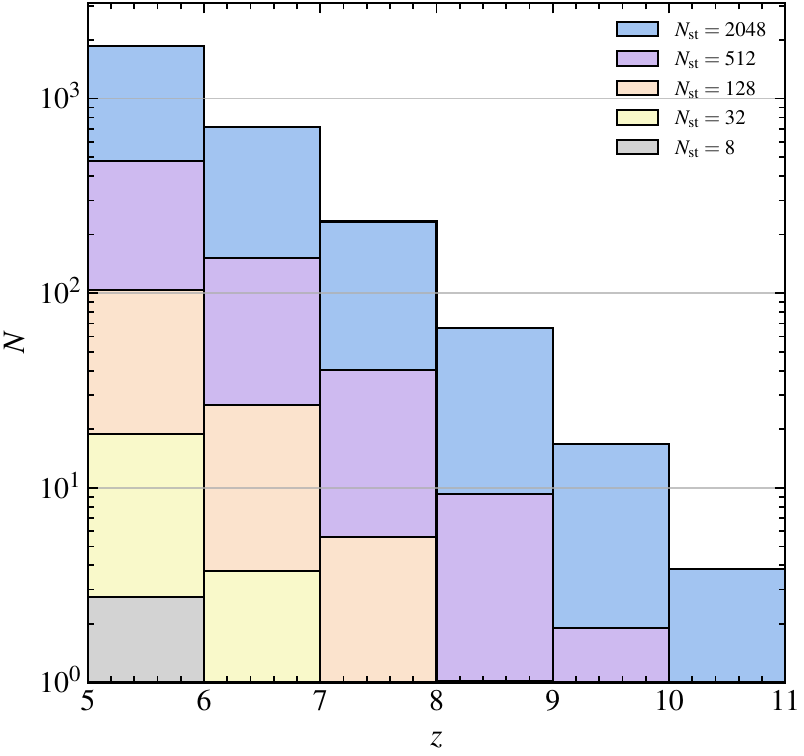}
    \caption{
    Similar to \reffg{fig:2}, but for 21-cm forest survey with $5\,{\rm kHz}$ frequency resolution.
    %The figure shows the variation of the number of observable quasars with redshift in the SKA-Low 21-cm forest survey in different effective reception areas.
    }\label{fig:3}
    \end{minipage}\hfill
%\end{figure}
%\begin{figure}
%    \centering
    \begin{minipage}[t]{0.48\textwidth}
    \includegraphics[width=\textwidth]{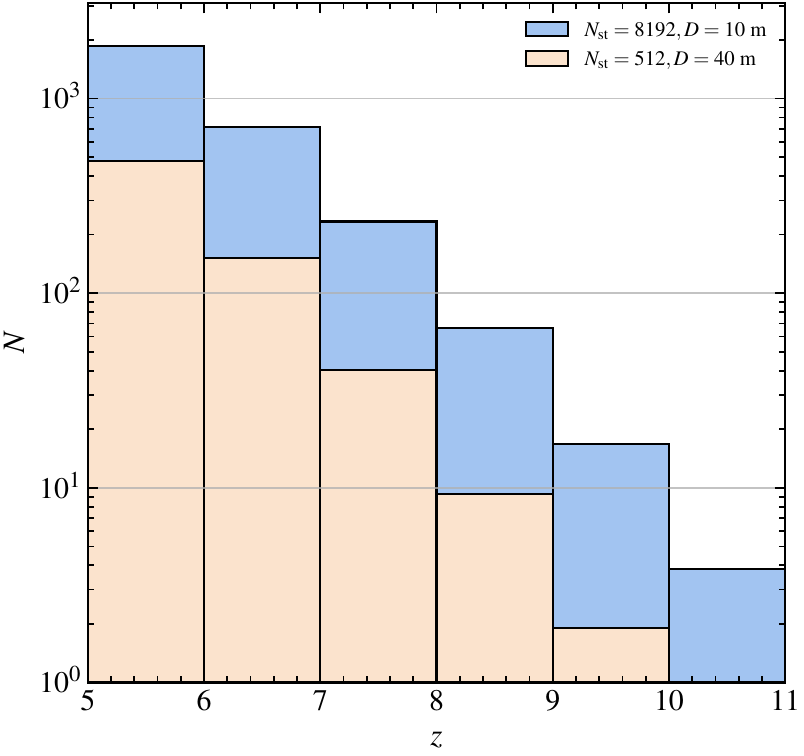}
    \caption{
    Similar to \reffg{fig:4}, but for 21-cm forest survey with $5\,{\rm kHz}$ frequency resolution.
    %The figure shows the number of quasars that can be observed in the 21-cm survey with the same deployment as described above. The results are not affected by changes in antenna length and the number of base stations. Therefore, the number of quasars obtained by the 21-cm survey is constant for the same effective receptive area
    }\label{fig:5}
    \end{minipage}
\end{figure*}

%The number of HzRLQs detected in 21-cm forest survey with  redshift bin equal to 1
%is shown in \reffg{fig:3}. 
%The light yellow histogram corresponding to an array of thirty-two $40\,{\rm m}$-diameter stations, 
%with an effective collecting area of $26187.5{\rm m}^2$ enables the detection of more than $10$ radio-loud quasars at $5<z<6$.
%With the increase in the number of stations, the most ambitious configuration ($N_{\rm st}= 2048$) achieves nearly 1000 detections at $z = 5$, while maintaining statistically robust samples ($N > 10$) throughout the entire redshift range up to $z = 10$.

Through wide-area continuum surveys, large low-frequency radio interferometric arrays have the potential 
to detect a substantial number of high-redshift radio-loud quasars (HzRLQs). 
By integrating over frequency channels, continuum observations significantly enhance the signal-to-noise ratio.
However, this integration comes at the expense of frequency resolution, 
which is critical for 21-cm forest power spectrum analysis. 
As a result, continuum detections often require dedicated follow-up observations
to probe fine spectral structures. 
In this work, we assume a frequency resolution of $5\,{\rm kHz}$ for 21-cm forest surveys, 
and investigate how different array sizes impact the detectability of HzRLQs. 
The corresponding results are shown in \reffg{fig:2}, 
where different colors represent low-frequency arrays of various sizes. 
For this analysis, we adopt a redshift bin width of $\Delta z = 1$.

Compared to continuum observations, the high spectral resolution required for 21-cm forest studies
leads to a significant increase in observational noise, thereby reducing the detectability of HzRLQs. 
For a mini-size array with only $n = 8$ stations, detecting HzRLQs beyond $z > 5$ becomes nearly impossible. 
However, for an array comparable to the ground-based SKA, with $n = 512$ stations, 
a moderate number of HzRLQs can still be detected, even though sensitivity remains lower than in continuum mode. 
In this configuration, the detection limit extends to approximately $z \sim 10$, 
which is sufficient for meaningful cosmological studies using the 21-cm forest. 
For extremely large arrays with $n = 2048$ stations, the detection redshift limit can be 
further extended to $z \sim 11$.

We also explore a scenario in which the total effective collecting area is constant, 
but the number of stations is increased while reducing the diameter of each station. 
This approach enhances the field of view and survey efficiency, potentially improving the HzRLQ detection rate. 
The results are presented in \reffg{fig:5}, where different colors indicate different configurations 
of station number and size. 
Our findings show that reducing the station diameter by a factor of four
and simultaneously increasing the number of stations by a factor of sixteen can substantially
increase both the number of detectable HzRLQs and the maximum redshift reached. 
However, this gain in survey efficiency comes with a trade-off: significantly greater data transmission, 
storage, and processing demands, posing major challenges for interferometric data analysis.

The detectability of HzRLQs in 21-cm forest surveys is governed by a complex interplay 
between array configuration and spectral resolution. 
While continuum surveys benefit from lower noise levels due to frequency integration, 
21-cm forest observations require high spectral resolution and thus face higher noise. 
Larger and denser interferometric arrays can extend detection capabilities to higher redshifts, 
but they also introduce considerable computational and infrastructural challenges. 
Achieving an optimal balance between sensitivity, resolution, and data processing 
efficiency will be essential for the successful design and operation of future 
low-frequency radio arrays targeting the 21-cm forest.

\subsection{onfusion limit}

\begin{figure}
    \centering
    \includegraphics[width=0.48\textwidth]{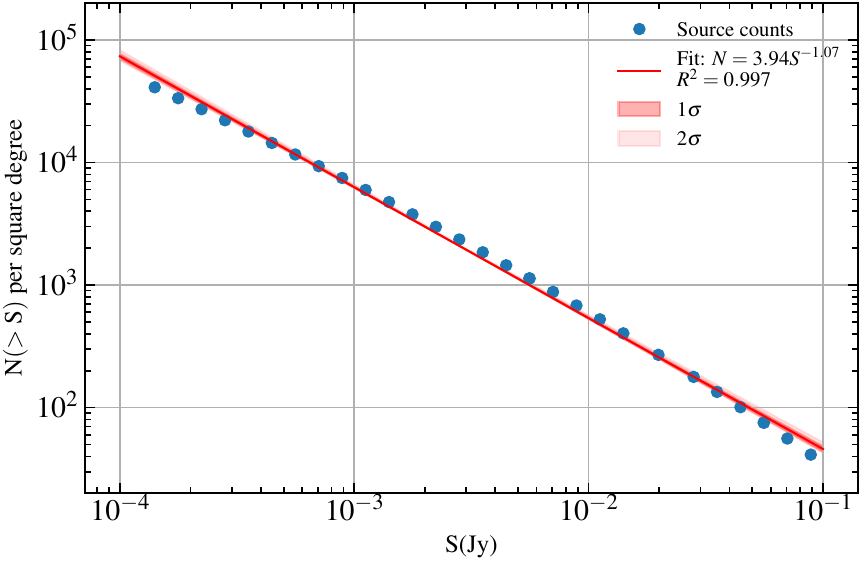}
    \caption{Power-law fitting of source counts at $5.5\lesssim z\lesssim 20.5$  and flux density at $200\rm MHz$ based on SKA-Low.}
    \label{fig:6}
\end{figure}

\begin{figure}
    \centering
    \includegraphics[width=0.48\textwidth]{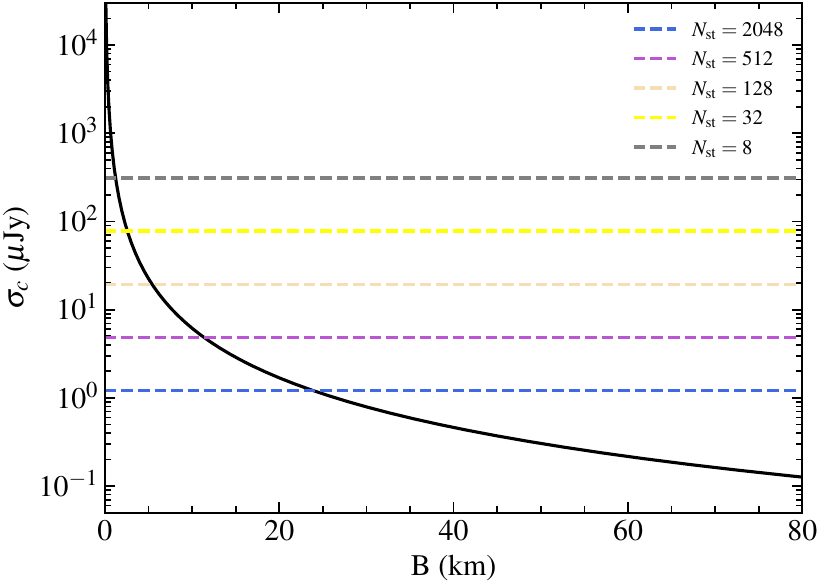}
    \caption{The black line is the curve of the confusion limit $\sigma_c$ as a function of the baseline length $B$. Different color lines correspond to the observational noise under different configuration schemes.}
    \label{fig:7}
\end{figure}

The confusion limit is reached when the density of sources brighter than the observational noise becomes high enough within the area of the synthesized beam. At this point, the ability of the radio array to resolve individual sources is limited not by sensitivity, but by resolution. Consequently, the baseline length of the interferometric array plays a critical role in determining the array’s effectiveness in mitigating source confusion.
With the luminosity function constructed in \refsc{sec:2}, 
we simulate the radio source samples with flux density ranging from 
$10^{-4} \sim 10^{-1} {\rm Jy}$, 
which is large enough to cover the faint end of the radio sources,  
within the redshift of $5.5 < z < 20.5$, and estimate the 
column number density as a function of flux density. 
We fit the flux-density cumulative column number density with a power law function,
%We derive by integration that the flux density ($S$) of the radio source obeys a power law distribution with respect to the number per square degree ($N$)
\begin{equation}
  N(>S)=CS^\beta,
\end{equation}
where $N(>S)$ represents the column number density with sources' flux density
larger than $S$, and a power law function is characterized by 
the free parameters $C$ and $\beta$.
The best-fit values are $C=3.94$, $\beta=-1.07$. 
The simulated flux-density cumulative column number density, as well as the
best-fit curve, are shown in \reffg{fig:6}.
%This fit is meant to describe the brightness of high-redshift quasars, as shown in 

The confusion limit is then estimated via \citep{2006MNRAS.369..281J,A.Cohen2004},
\begin{equation}
    \sigma_c=(\theta^2mC)^{-1/\beta},
\end{equation}
where $C=3.94$, $\beta=-1.07$ are the parameters of the best-fit power law function,
$m = 30$ is the threshold parameter \citep{2001AJ....121.1207H}, and 
%We use the computation of the confusion limit in  where 30 sources form a beam  . 
$\theta$ is the angular resolution. 
The angular resolution of the interferometer array is related to the 
maximum length of the baseline, e.g. $\theta = c/(\nu B_{\rm max})$, where
$c$ is the speed of light and $\nu$ is the corresponding frequency.
To simplify the estimation, we ignore the frequency dependency and use 
$\nu=200\,{\rm MHz}$ in the following analysis. 
%equal to $c/\nu B$ and we take the center frequency of SKA-Low, $\nu=200 \rm MHz$, as our detection source frequency.

We show the confusion limit as a function of the maximum baseline lengths in
\reffg{fig:7}. The noise level with different numbers of $40\,{\rm m}$ stations
is shown with horizontal dashed lines in different colors.
It is evident that increasing the maximum baseline length significantly reduces the confusion limit. To ensure that the confusion limit remains below the system’s thermal noise level, a minimum threshold for the maximum baseline should be established for configurations with varying numbers of stations. In general, a maximum baseline length of at least approximately $25\,{\rm km}$ is sufficient to render the confusion limit negligible.

%\red{We can get that the longer the baseline, the smaller the confusion limit is, and the number of sources in the unit resolution cell exceeds the resolution capability of the telescope, leading to signal superposition. When the confusion limit is larger than the observational noise, the quality of the observed data is mainly limited by the spatial resolving capability. In the several configurations we considered, the longest baseline length when the confusion limit is the dominant factor is $\sim 23.8 \rm km$ while the shortest baseline case also needs to be constructed at $\sim 1.2 \rm km$. It is worth noting that the interferometric arrays with a shorter baseline are generally constructed on the Moon with a better ability to ignore the observational noise compared to the baseline length of the Earth for SKA-Low.}

%\red{To reduce the impact of the confusion limit on the observation, we can still adopt new strategies in data processing, such as using machine learning algorithms and other methods to de-confuse the limit and isolate the source of confusion from the image to obtain more accurate observations, while the array configuration remains unchanged\citep{2020PhRvD.102h3024B}.}

\subsection{Challenges in construction}

As we discussed in the previous sections, the substantial increase in the number of base stations
brings a series of complex challenges in hardware design, data processing, and data storage \citep{2008Ap&SS.313....1T}.

The construction of large-scale radio arrays faces significant technical challenges 
as the number of antenna elements increases. Each station comprises multiple critical components, 
including antennas, low-noise amplifiers, filters, and analog-to-digital converters, 
With the dramatic rise in real-time data throughput, network congestion becomes a serious concern, 
particularly in distributed or remote configurations \citep{wang2015squarekilometerarraytelescope}. 

These challenges are further amplified for space-based applications such as lunar interferometer arrays, 
where observational data must be downlinked to ground-based data centers for further analysis. 
The resulting data volumes present substantial hurdles for storage, transport, and long-term data management.

From a computational perspective, system scaling results in the combinatorial growth of processing complexity. 
The calibration and imaging pipelines generate intermediate data products that often exceed the volume
of raw observational data by orders of magnitude \citep{2019SCPMA..6289531A,2024A&A...692A..61W}.
%For international projects like the SKA, additional constraints emerge from the requirement for global data sharing, which imposes strict demands on metadata standardization (e.g., IVOA protocols) and sophisticated data retrieval systems \citep{Pritchard_2025}. The long-term preservation of radio astronomical data for reproducible research presents another critical challenge, requiring periodic data migration to prevent format obsolescence that incurs substantial ongoing costs.

Given that the data transfer rate of SKA-Low is as high as $7.15\, \rm Tbps$ \citep{2023SSPMA..53v9504G} and the size of the data volume grows exponentially with the number of baselines ($N_{\rm st}(N_{\rm st}-1)/2$), such a large amount of data makes it technically challenging to transmit raw observation data from the Moon-based interferometer directly back to the Earth. 
To address the challenges of data transmission and real-time data processing in Moon exploration, we propose to establish a dedicated Moon-based data processing center. This facility will enable in situ data analysis, significantly reducing reliance on Earth-based processing while optimizing the science return from lunar missions.

Furthermore, dense station layouts heighten the risk of self-generated radio frequency interference (RFI), 
which can significantly degrade data quality. Mitigating these effects requires meticulous array design, 
including optimized station geometries, careful component isolation, 
and advanced electromagnetic shielding techniques \citep{2016RAA....16...36H}.

Overall, these technical and infrastructural challenges must be carefully considered 
in the design of next-generation radio arrays, particularly those targeting high-resolution, 
wide-field observations such as the 21-cm forest.

\section{Conclusion} \label{sec:4}

This work investigates the scientific prospects and technical design requirements for detecting 
high-redshift radio-loud quasars (HzRLQs) using a Moon-based low-frequency interferometric array, 
with the goal of enabling 21-cm forest studies during the Epoch of Reionization (EoR) and Cosmic Dawn.

We begin by improving the quasar luminosity function (QLF) model through the inclusion of dust obscuration effects. Since optical surveys tend to miss heavily obscured sources, particularly at the faint end, we introduce an X-ray-based correction to account for this bias. The obscuration-corrected QLF yields more accurate predictions for the abundance of HzRLQs and helps partially mitigate discrepancies between theoretical models and observations.

Assuming a total survey area of $10^4\,\mathrm{deg}^2$ and an integration time of 1 year, we analyze the detectability of HzRLQs in two complementary observing modes: continuum surveys and 21-cm forest surveys. For continuum surveys, we adopt a bandwidth of $10\, {\rm MHz}$, enabling significant signal-to-noise enhancement through frequency integration. In contrast, 21-cm forest observations require fine frequency resolution
-- assumed to be $5\,{\rm kHz}$ in this work -- to resolve small-scale absorption features, which increases thermal noise and reduces source detectability.

We evaluate detection performance across a range of array sizes. For the smallest configuration considered ($N_\mathrm{st} = 8$ stations with $40\,{\rm m}$ diameter), continuum surveys can detect HzRLQs up to $z \sim 10$, but the 21-cm forest survey becomes ineffective beyond $z \sim 5$. For SKA-scale arrays ($N_\mathrm{st} = 512$), continuum surveys can detect quasars out to $z \sim 16$, while 21-cm forest observations remain feasible up to $z \sim 10$. For an extremely large array with $N_\mathrm{st} = 2048$, the detection limit of the 21-cm forest survey extends to $z \sim 11$, supporting deeper exploration of the EoR.
To minimize the confusion effects, the maximum baseline length needs to be $\gtrsim 25\,{\rm km}$.

We further explore configurations that maintain a fixed total collecting area ($\sim 419000\,\mathrm{m}^2$) but increase the number of stations by reducing the station diameter. This strategy improves field-of-view and survey efficiency, enabling higher detection rates without additional collecting areas. For example, replacing the baseline configuration ($N_\mathrm{st}=512$, $D_\mathrm{st}=40$\,m) with $(N_\mathrm{st}=8192, D_\mathrm{st}=10\,\mathrm{m})$ substantially increases the number of detectable HzRLQs and raises the detection redshift limit. However, this improvement comes with greater demands on real-time data transmission, onboard storage, calibration, and computing infrastructure, particularly challenging for a Moon-based observatory.

In summary, the detectability of HzRLQs depends sensitively on array configuration, frequency resolution, and survey design. Continuum surveys provide broad coverage with high signal-to-noise, while 21-cm forest observations offer detailed spectral information at the cost of sensitivity. Both modes benefit from careful optimization of array layout and observing parameters. Future efforts should focus on improving models of high-redshift quasar obscuration, refining array deployment strategies, and developing scalable data processing pipelines. With these advancements, a lunar low-frequency interferometer could open a powerful new observational window into the reionization era, enabling high-precision studies of structure formation, dark matter, and the thermal history of the early Universe.

\section*{Acknowledgments}

This research was supported by multiple funding sources: the National SKA Program of China (2022SKA0110200, 2022SKA0110203), the National Natural Science Foundation of China (12473091, 12473001), and the National 111 Project (B16009). 
YL acknowledges the support of the Fundamental Research Funds for the Central Universities (No. N2405008).
We express our gratitude to ChatGPT for its assistance in refining the manuscript and enhancing the code to improve execution efficiency.

\bibliography{ref}

%% This command is needed to show the entire author+affiliation list when
%% the collaboration and author truncation commands are used.  It has to
%% go at the end of the manuscript.
%\allauthors

%% Include this line if you are using the \added, \replaced, \deleted
%% commands to see a summary list of all changes at the end of the article.
%\listofchanges
%\end{CJK*}
\end{document}